\documentclass[intlimits,twoside,a4paper]{article}

\usepackage{amsmath,amssymb}
\usepackage{graphicx}

\usepackage[T2A]{fontenc}
\usepackage[cp1251]{inputenc}

\usepackage[eqsecnum]{cmpj2}

\issue{2016}{19}{1}{13602}
\doinumber{10.5488/CMP.19.13602}

\title[Anionic and nonionic surfactants at the SiO$_{2}$/water interface]
{Computational studies on the behaviour of anionic and nonionic
surfactants at the SiO$_{2}$ (silicon dioxide)/water interface}

\author[E. N\'u\~nez-Rojas, H. Dominguez]{E. N\'u\~nez-Rojas\refaddr{label1},
H. Dominguez\refaddr{label2}\footnote{%
    Corresponding author. E-mail: hectordc@unam.mx.
    Present address: Department of Physics
    and Astronomy, University of British Columbia, Vancouver, British
    Columbia, Canada.  On sabbatical leave.}}
\addresses{
\addr{label1} Departamento de Qu\'imica, Universidad Aut\'onoma
Metropolitana-Iztapalapa, M\'exico, D.F. 09340
\addr{label2} Instituto de Investigaciones en Materiales,
Universidad Nacional Aut\'onoma de M\'exico,
M\'exico, D.F. 04510
}

\date{Received October 15, 2015, in final form November 23, 2015}
\authorcopyright{E. N\'u\~nez-Rojas, H. Dominguez, 2016}

\begin{document}
\maketitle

\begin{abstract}

Molecular dynamics simulations to study the behaviour of
anionic (Sodium Dodecylsulfate, SDS) and nonionic (Monooleate of Sorbitan,
SPAN80) surfactants close to a SiO$_{2}$ (silicon dioxide) surface were
carried out. Simulations showed that a water layer was first
adsorbed on the surface and then the surfactants were attached on that
layer. Moreover, it was observed that
water behaviour close to the surface influenced the
surfactant adsorption since a semi-spherical micelle was formed
on the SiO$_{2}$ surface with SDS molecules
whereas a cylindrical micelle was formed with SPAN80 molecules.
Adsorption of the micelles was
conducted in terms of structural properties (density
profiles and angular distributions) and dynamical behaviour (diffusion
coefficients) of the systems.
Finally, it was also shown that some
water molecules moved inside the solid surface and located at
specific sites of the solid surface.


\pacs 68.08De, 68.43.Hm, 68.43.Jk
\end{abstract}

\section{Introduction}

Adsorption of surfactant molecules at solid-liquid
interfaces has been investigated for years not only for its relevance
in science but also for its numerous industrial applications, such as
detergency, crude oil refining, treatment of waste water, adsorption
on activated charcoal and even in pharmaceutical preparations
\cite{ah:sh,am:ja,PDD}.

In particular, self-assembly of surfactant molecules on solid
surfaces has shown different issues from
those observed at liquid/vapor and at liquid/liquid interfaces.
For instance, it has been observed that interactions between
hydrophobic tails,
repulsions between headgroups and interactions between surfactant molecules
with solid surface \cite{ga:fu,bo:ko} could
change the isotherms at the critic micellar concentration (CMC).
Therefore, studies of surfactant aggregation will help us to
obtain more physical
insights of self-assembly phenomena \cite{sa:pa,gu:xi}.

From the experimental point of view, the
surfactant adsorption on surfaces has been studied
by different techniques, such as streaming potential
methods \cite{fuers}, calorimetry \cite{ki:fi}, neutron
reflection \cite{pe:st}, ellipsometry \cite{ti:jo}, fluorescence
spectroscopy \cite{ch:so} and atomic force microscopy (AFM) \cite{du:wa}.
In fact, AFM has proved to be a reliable technique to obtain
information on the topology of surfactant aggregation since it allows us
to observe how surfactants are formed on surfaces. For instance,
CTAB arrays in parallel stripes on a graphite surface \cite{ma:cl},
SDS forms hemimicelles on a rough gold surface \cite{sc:sh} and similar
morphologies have been seen for other surfactants on hydrophobic
surfaces \cite{pa:wa,ja:bu,wa:du}.

On the other hand, computer simulations have been very useful
to study such complex systems. For instance, Monte Carlo simulations
have been used to provide information on
structural transitions of surfactant  aggregation \cite{zh:zh}
while molecular dynamics simulations have been used to investigate
aggregation at atomistic scales \cite{sh:ch,hec}.
In previous papers we have reported the surfactant behaviour
on different surfaces
\cite{hec,edg:hec,edg1:hec,car:hec}.
It has been observed how graphite surfaces impose orientational
order on the surfactant tails \cite{hec,car:hec} and how different solid faces
of a titanium dioxide produced different aggregates on the
surface \cite{edg:hec,edg1:hec}. In the present work we are
interested to extend the
studies of surfactant aggregation on a
hydrophilic SiO$_{2}$ (silicon dioxide) surface in order to compare the
behaviour of surfactant molecules on different solid substrates.

\section{Computational method and model}

\begin{figure}[!h]
\centerline{
\includegraphics[width=0.69\textwidth]{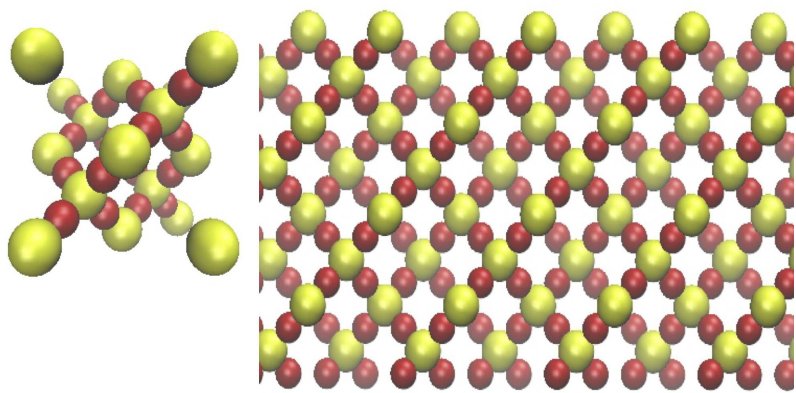}
}
\caption{(Color online) The unit crystal cell and a snapshot
of the solid surface. Oxygen atoms are in red and silicon atoms
are in yellow.}
\label{fig1}
\end{figure}

Molecular dynamics simulations of anionic and
nonionic surfactant molecules at silicon dioxide surface (SiO$_{2}$),
in its Cristobalite form, were carried out for the present study.
The surface was constructed using an atomistic model with a surface
orientation (001) (figure~\ref{fig1}). The parameters used for a solid
surface were taken from reference \cite{do:ga}.

For the anionic surfactant molecule, sodium dodecyl sulfate (SDS), a
model of 12 united carbon atoms
attached to a headgroup (SO\( _{4} \)) was simulated by using
a force field already reported in the literature \cite{hec}.
For the nonionic surfactant molecule, Monooleate of Sorbitan (SPAN80), there was used, a model
with 17 united carbon atoms, three OH$^{-}$ and one ester groups in the head
group.
The force field reported in reference \cite{edg1:hec} was used for this molecule.

In the case of simulations with the anionic molecules, the initial configuration
was prepared from a monolayer of 36 molecules in all-trans-configuration with
the headgroups initially pointed to the solid surface.
Then, 2535 water molecules were added (using the SPC model
\cite{be:gr}) to the system and 36 sodium cations (Na$^{+}$).
In the case of simulations with nonionic molecules,
25 molecules were used with the same number of water molecules.

A simulation box having dimensions $X = Y = 43.7019$
and $Z = 150$~{\AA} was used with the usual periodic boundary
conditions. The $Z$-dimension of the box was long
enough to prevent the formation of a second water/solid interface due to the
periodicity of the system. Instead, a liquid/vapor interface was present at one
end of the box ($z > 0$) whereas at the other end of the box
($z< 0$) beyond the solid there was an empty space.
All simulations were carried out in the NVT ensemble
with a time step of 0.002~ps using DL-POLY package \cite{fo:sm}. Bond lengths
were constrained using SHAKE algorithm with a tolerance of $10^{-4}$, and the
temperature was controlled using the Hoover-Nose thermostat having a relaxation
time of 0.2~ps \cite{hoov} at $T=298$~K.
Long-range electrostatic interactions were handled using
the Particle Mesh Ewald method, and the
Van der Waals interactions were cut off at 10~{\AA}. Finally,
the simulations were run up to 40~ns and
configurational energy was monitored as a function of time in order to determine the moment the systems have reached equilibrium. Then, the last 2~ns were collected
for analysis.

\clearpage

\section{Results}

In this section we present calculations of the surfactant
molecules at the SiO$_{2}$ surface. Studies on the behaviour of the surfactant
molecules and how they aggregate at the liquid/solid interface are discussed.

\subsection{Density profiles}

\begin{figure}[!b]
\centerline{
\includegraphics[width=0.45\textwidth]{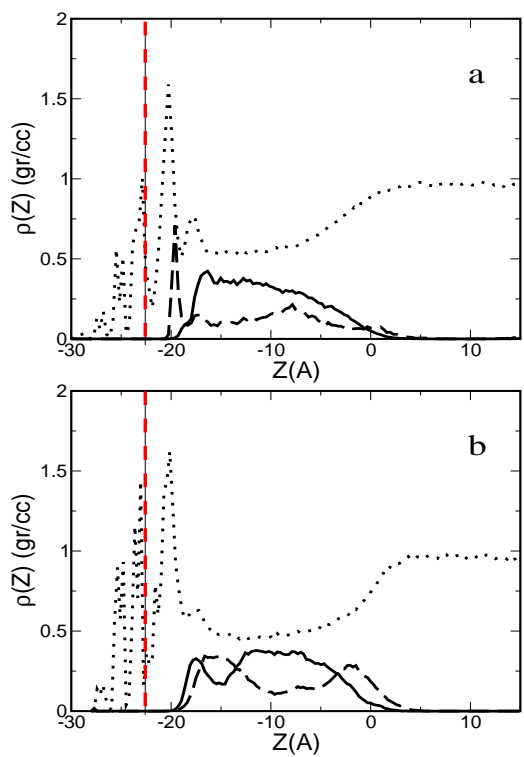}
}
\caption{Density profiles calculated along the
  $Z$-direction of the SiO$_{2}$ solid surface. (a) SDS surfactant
  and (b) SPAN80 surfactant. Dotted lines represent water, dashed
  lines show a surfactant headgroup and continuous lines show surfactant tails.
  The solid surface position is indicated by the red line.}
\label{fig2}
\end{figure}

In order to determine where the surfactant molecules arrayed in the system,
mass $Z$-dependent density profiles for the headgroups and the tails were
calculated, i.e., normal to the liquid/solid interface.

From figure~\ref{fig2} we observed that water molecules (dotted line) were not
only adsorbed but also absorbed by the solid surface which
is located at a position of $Z = -23$~{\AA} in the figure. In fact,
the first water profile peak (to the right of the surface)
indicated strong adsorption, i.e., a water layer on the surface.
The other water peaks (to the left of the surface) suggested that few particles
were inside the solid surface. The presence of water molecules
inside SiO$_{2}$ surfaces has been also observed in real
experiments \cite{anto,ame}. On the other hand,
SDS density profiles showed a strong first peak for
the polar group ($\approx 4$~{\AA} from
the surface in figure~\ref{fig2}~(a)
suggesting that the surfactant was well adsorbed on the surface. Moreover,
it was noted that the peak was located to the right of the
adsorbed water layer. The result showed that the
surfactant molecules formed a micellar structure adsorbed on the water layer
on the solid surface. It was also possible to
observe that the hydrocarbon chains (solid line) were sited between
polar-groups (dashed line) along $Z$ direction.

In the case of simulations with SPAN80 molecules [figure~\ref{fig2}~(b)],
the water density  profile also showed that some of those  molecules
went deep into the substrate.
Besides, unlike SDS system, here strong
peaks were observed for the water profiles. The SPAN80 profiles are also shown,
the first peak ($\approx 5$~{\AA} from the
surface) corresponded to the hydrocarbon
chains and it was very close to the polar group, nevertheless
the hydrocarbon chains were
surrounded by the polar groups. These profiles suggested the
formation of a micelle on the surface.

\begin{figure}[!t]
\centerline{
\includegraphics[width=0.9\textwidth]{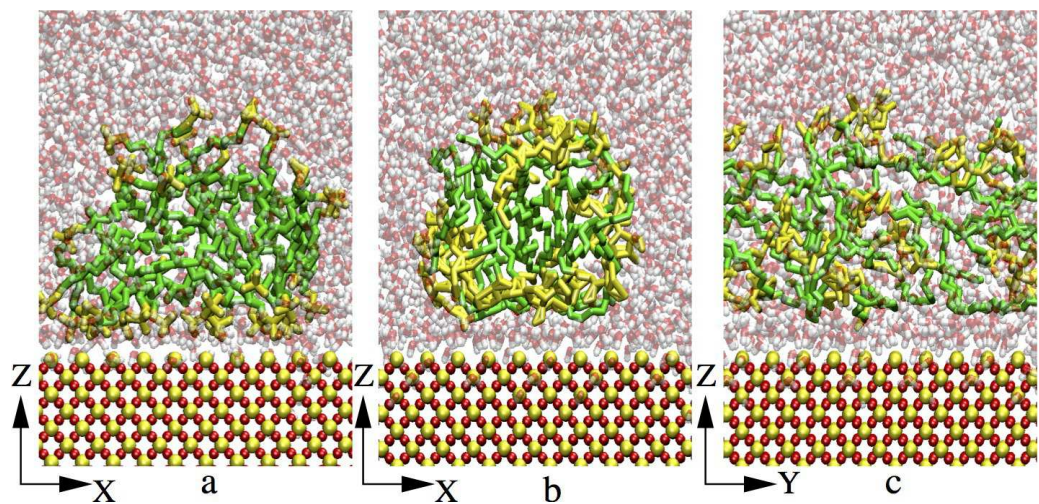}
}
\caption{(Color online) Snapshots for SDS (a) and for SPAN80 (b and c)
  micelles on a SiO$_{2}$ surface.
  Spherical and cylindrical shapes are depicted
  for the SDS and SPAN80, respectively. Tail groups are in green and
head groups in yellow.}
\label{fig3}
\end{figure}

\begin{figure}[!b]
\centerline{
\includegraphics[width=0.8\textwidth]{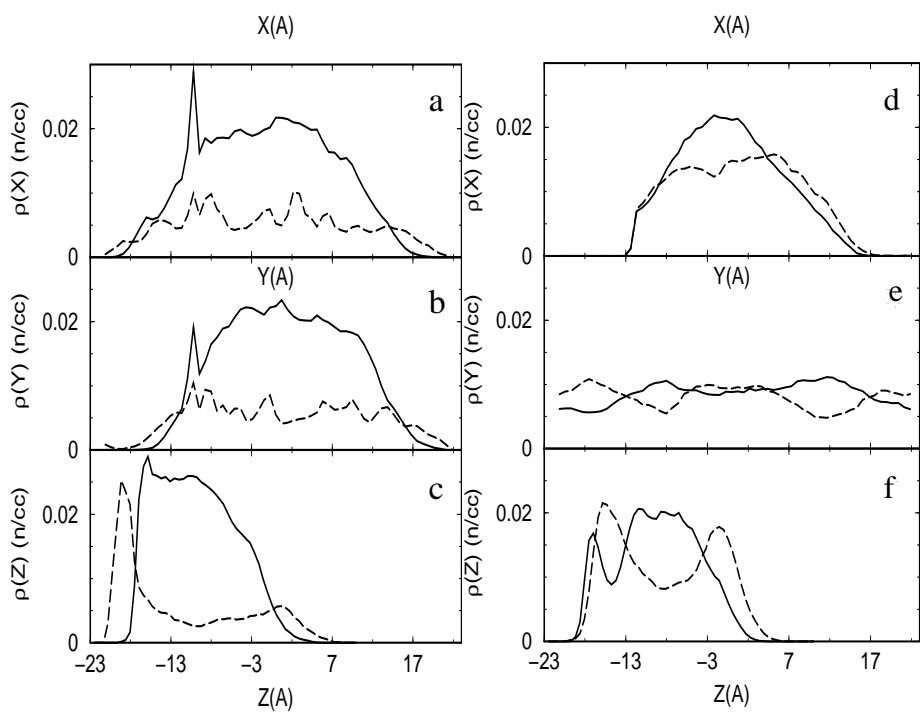}
}
\caption{Particle density profiles of the
  surfactant headgroups (dashed lines) and tails (solid lines)
  for the SDS (left-hand) and SPAN80 (right-hand) systems. The panels indicate
  the profiles in the $X$, $Y$ and $Z$ axis.}
\label{fig4}
\end{figure}

In figure~\ref{fig3}, snapshots of the final SDS [figure~\ref{fig3}~(a)] and SPAN80 [figures~\ref{fig3}~(b) and \ref{fig3}~(c)] are shown. In those figures it was possible to observe the micellar
structure mentioned above. In the case of the SDS, the micelle
structure had a spherical-like shape (snapshot in $X-Z$ and $Y-Z$ looked alike)
whereas for SPAN80 the micelle
had a cylindrical-like structure [see figures~\ref{fig3}~(b) and \ref{fig3}~(c)].

In order to verify the structure of the micelles on the surfaces,
particle density profiles in the other two directions, $X$ and $Y$, were calculated.
In figures~\ref{fig4}~(a) and \ref{fig4}~(b), head groups density profiles
(dashed line) and tail groups (solid line) of SDS
molecules are shown. There were observed
similarities between the two particle profiles, $\rho(X)$ and $\rho(Y)$,
where it was noted that polar groups surrounded
the tail groups.
In both tail group profiles, a small sharp peak can be seen.
In this case, SDS molecules are well adsorbed on the SiO$_{2}$ substrate
with some surfactants attached on the surface by the head
group [see figure~\ref{fig3}~(a)].
Then, there is a possibility that a SDS molecule could remain anchored with its
tail moving in one region only. This could explain the small sharp peak
observed in both profiles, i.e.,
there are a bit more tail groups in that region.
On the other hand, from the $Z$-particle
density profile in figure~\ref{fig4}~(c), a big head group peak could be seen close to the
solid surface (see also figure~\ref{fig2}), i.e., there was an
excess of head groups attached to the surface. Moreover, the
$\rho(Z)$ indicated that the size of the micelle in the $Z$-direction
was smaller than in the $X-Y$ directions. Therefore,
the structure can be described as a sphere deformed along the
perpendicular direction of the solid surface.

This calculation was also carried out for the
SPAN80 molecules [figures~\ref{fig4}~(d)--\ref{fig4}~(f)]. In figure~\ref{fig4}~(d), the density profile along $X$-axis
is shown where a molecular aggregation is observed. In figure~\ref{fig4}~(e), the profiles indicated a uniform distribution of the molecules along
the $Y$-axis whereas in figure~\ref{fig4}~(f) it was again possible to observe
aggregation of the molecules next to the surface.
Here, it can be seen that headgroups partially surrounded the
tail groups. Then, these profiles revealed that the
molecules structured themselves as a cylindrical micelle along
the $Y$ direction. These results suggested
that the effect of the solid surface in the micelle formation
was minimum.

\subsection{Water orientation at the SiO$_{2}$ surface}

As it was observed above, water density profiles showed strong peaks
close to the SiO$_{2}$ surface suggesting that those molecules
might have some structure close to the solid.
Therefore, studies of how water molecules were oriented in the solid
were carried out. The analyses were conducted
over the molecules in the adsorbed layers only (defined by the peaks of figure~\ref{fig2}).

\begin{figure}[!h]
\centerline{
\includegraphics[width=0.98\textwidth]{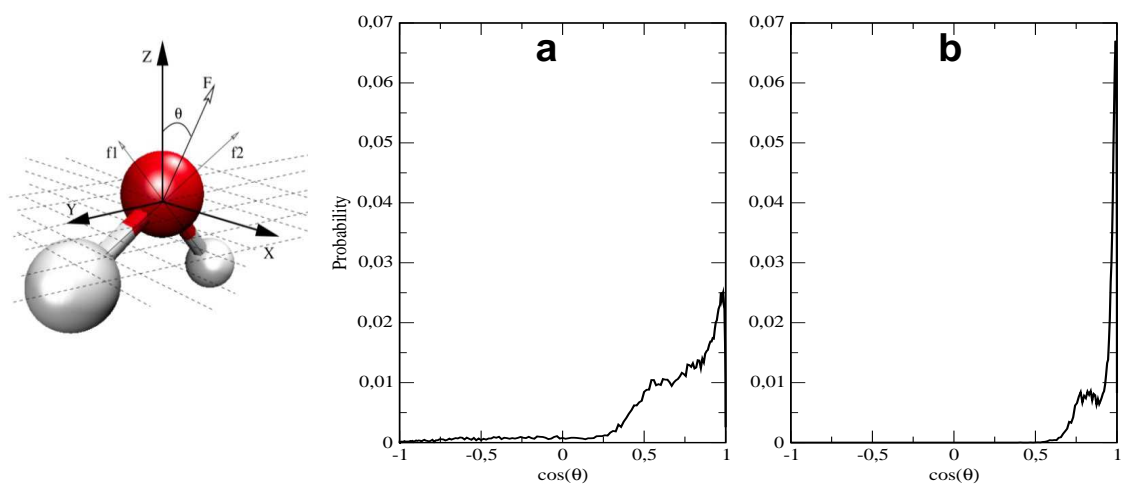}
}
\caption{(Color online) Angular probability of the water dipole
vector with the vector normal to the interface. (a) For the SDS
and (b) for the  SPAN80 surfactant systems.}
\label{fig5}
\end{figure}

In figure~\ref{fig5}, angular distributions of
water molecules are shown for the SDS and SPAN80 systems. The orientation
of the water molecules was measured as the angle between the bisector
vector of the OH bonds and the vector normal to the interface (see
figure~\ref{fig5}). In both systems, SDS and SPAN80, there was observed a privileged
orientation of the water molecules, i.e., they were nearly perpendicular
to the surface. In fact, there was noted a larger number of water molecules
pointing perpendicular to the surface
for the SPAN80 system [high peak
in figure~\ref{fig5}~(b)] than for the SDS system (small peak). Moreover,
the water angle distribution presented a small peak when water
interacted with SPAN80 [figure~\ref{fig5}~(b)] and a wider distribution
when water interacted with SDS.

\begin{figure}[!t]
\centerline{
\includegraphics[width=0.7\textwidth]{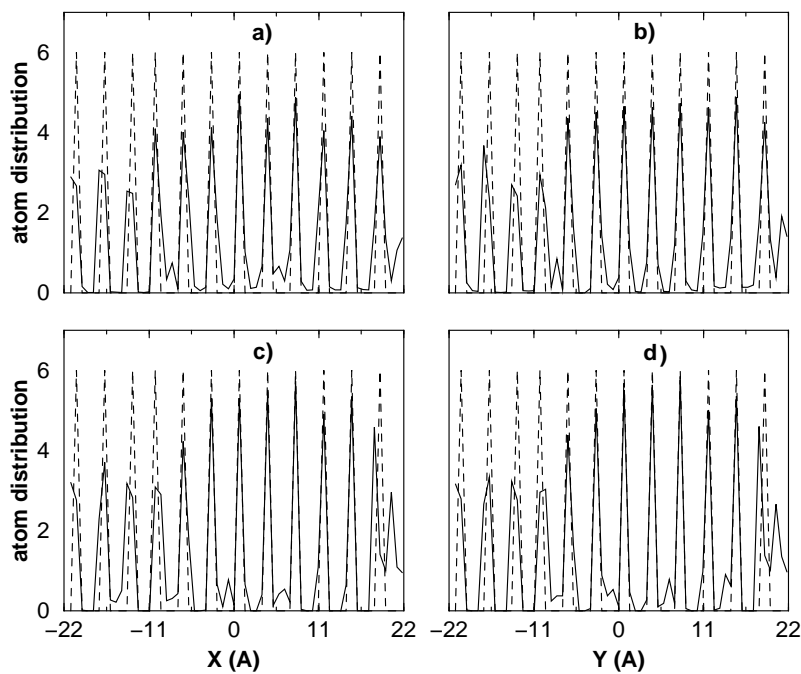}
}
\caption{Distribution of oxygen atoms in
  water (solid lines)
  and silicon  atoms of the SiO$_{2}$ surface (dashed lines).
  Panels (a) and (b) atom positions for the SDS surfactant in the $X$ and $Y$ axis.
  Panels (c) and (d) atom positions for the SPAN80 surfactant
  in the $X$ and $Y$ axis.}
\label{fig6}
\end{figure}

In figure~\ref{fig6}, the distribution (in the first adsorbed water layer)
of water molecules in the $X$ and $Y$ directions, i.e., in the $X-Y$ plane,
is plotted. Figures~\ref{fig6}~(a) and \ref{fig6}~(b) show the water oxygens positions
in the solid surface for the SDS system.
In the same figure~\ref{fig6},
the positions of the silicon atoms of the
SiO$_{2}$ surface are also shown in the same $X-Y$ directions.
In all cases it was observed that
water molecules (solid lines) were located
only in specific sites.
In fact, they were placed on the silicon atoms of the solid surface
(indicated by the dashed lines in figure~\ref{fig6}).
Figures~\ref{fig6}~(c) and \ref{fig6}~(d) are the same plots for
the SPAN80 system. From the above results we noted that water oxygens were
adsorbed above silicon sites.

\subsection{Mobility of surfactant molecules}

In order to describe the attachment of the surfactant molecules to
the solid surface, the diffusion coefficient of the
molecular aggregates was also calculated. The diffusion coefficients were calculated
in each direction by measuring the square mean displacements
of the surfactant atoms and using the Einstein
relation. For the SPAN80 surfactant, the diffusions were
$0.126 \times 10^{-9}$~m$^2$/s, $0.095 \times 10^{-9}$~m$^2$/s
and $0.085 \times 10^{-9}$~m$^2$/s in the
$X$, $Y$ and $Z$ directions, respectively. For the SDS surfactant,
the diffusions were $0.125 \times 10^{-9}$~m$^2$/s, $0.135 \times 10^{-9}$~m$^2$/s
and $0.078 \times 10^{-9}$~m$^2$/s in the $X$, $Y$ and $Z$ directions, respectively.
In both systems, the diffusions in the perpendicular direction ($Z$-axis)
was lower than in the plane. However, the values for the SPAN80
were higher than for the SDS surfactant, suggesting that the
second surfactant had a higher mobility than the first one.
It is worthy to mention that these diffusions were much lower
than the diffusions found for the same molecules on a
Titanium dioxide surface \cite{edg:hec,edg1:hec}.

\section{Conclusions and discussion}

A series of Molecular Dynamics simulations were carried out in order to
describe the behaviour of two different surfactant molecules interacting
with a silicon dioxide solid surface. In the case of
the anionic surfactant (SDS), there was observed
a spherical micelle formation on a layer of water molecules
previously adsorbed on the solid. The micelle was described by density
profiles and they showed a deformation of the micelle
next to the adsorbed layer of water molecules.
This deformation can be explained in terms of
the SDS charged headgroups interactions
with water molecules on the solid surface.
On the other hand, the nonionic surfactant (SPAN80)
did not show much influence by the solid surface. It
formed a cilyndrical micelle next to the adsorbed water layer.
In this case, there was observed a thicker water layer
between the surfactant and the solid surface.
The influence of the surfactant on the surface was characterized
by water molecules in the surface. Dipole water orientation,
in the solid surface, was more tilted for the SDS molecules than for the
SPAN80 suggesting a stronger SDS-surface interaction and consequently
more intensive adsorption of those molecules on the surface.
Based on the previous results (diffusion coefficients),
it was noted that both surfactants showed less affinity
with the SiO$_{2}$ surface than with other surfaces
such as graphite and titanium dioxide.

\section*{Acknowledgements}

HD acknowledges support from DGAPA-UNAM-Mexico and
Conacyt-Mexico for sabbatical scholarships. The authors acknowledge support from DGTIC-UNAM for
 the supercomputer facilities.


\clearpage

\ukrainianpart

\title{Чисельні дослідження поведінки аніонних та неіонних сурфактантів на границі розділу між двоокисом кремнію  (SiO$_{2}$) і водою}

\author{E. Нуньєс-Рохас\refaddr{label1},
Г. Домінгес\refaddr{label2}}
\addresses{
\addr{label1} Хімічний факультет, Автономний університет Метрополітана-Іцтапалапа, Мехіко, Мексика
\addr{label2} Інститут матеріалознавства, Національний автономний університет Мексики, Мехіко, Мексика
}

\makeukrtitle

\begin{abstract}
Виконано моделювання поведінки аніонних (додецилсульфат натрію, SDS) та неіонних (монолеат сорбітану, SPAN80) сурфактантів поблизу поверхні  SiO$_{2}$ (двооксиду кремнію) методом молекулярної динаміки. Спостерігаємо, що шар води спочатку адсорбується на поверхні, а тоді сурфактанти приєднуються до цього шару. Більш того, спостережено, що поведінка води поблизу поверхні впливає на адсорбцію суфрактантів, оскільки напів-сферична міцела  утворюється на поверхні  SiO$_{2}$ з SDS молекулами, в той час, як циліндрична міцела формується у випадку молекул  SPAN80. Адсорбція міцел описана структурними властивостями (профілі густини і кутові розподіли), а також і в термінах коефіцієнтів дифузії. Показано, що деякі молекули води проникають всередину твердої поверхні.

\keywords комп'ютерне моделювання, SDS сурфактант, SPAN80 сурфактант, адсорбція, кристобаліт

\end{abstract}

\end{document}